\documentclass[pra, amsmath,amssymb,aps,10pt,twocolumn,superscriptaddress]{revtex4-1}
\usepackage{graphicx}% Include figure files
\usepackage[usenames, dvipsnames]{color}
\usepackage{dcolumn}% Align table columns on decimal point
\usepackage{bm}% bold math
\usepackage{color}
\usepackage[colorlinks,bookmarks=false,citecolor=blue,linkcolor=blue,urlcolor=black]{hyperref}

\begin{document}

\preprint{APS/123-QED}

\title{Strong-field triple ionization of atoms with $p^3$ valence shell}
\author{Jakub S. Prauzner-Bechcicki}\email{jakub.prauzner-bechcicki@uj.edu.pl}
\affiliation{Instytut Fizyki imienia Mariana Smoluchowskiego, Jagiellonian University in Krakow, \L{}ojasiewicza 11, 30-348 Krak\'ow, Poland}

\author{Dmitry K. Efimov} 
\affiliation{Institute of Theoretical Physics, Jagiellonian University in Krakow, \L{}ojasiewicza 11, 30-348 Krak\'ow, Poland}
\affiliation{Department of Theoretical Physics, Faculty of Fundamental Problems of Technology, Wroc\l{}aw University of Science and Technology, 50-370 Wroc\l{}aw, Poland}
\author{Micha\l{} Mandrysz}
\affiliation{Institute of Theoretical Physics, Jagiellonian University in Krakow, \L{}ojasiewicza 11, 30-348 Krak\'ow, Poland}
\author{Jakub Zakrzewski}
\affiliation{Institute of Theoretical Physics, Jagiellonian University in Krakow, \L{}ojasiewicza 11, 30-348 Krak\'ow, Poland}
\affiliation{Mark Kac Complex Systems Research Center, Jagiellonian University in Krakow, \L{}ojasiewicza 11, 30-348 Krak\'ow, Poland}
\date{\today}% It is always \today, today,
%  but any date may be explicitly specified
 
\begin{abstract}
The interaction of strong pulsed femtosecond laser field with atoms having three equivalent electrons in the outer shell ($p^3$ configuration, e.g. nitrogen) is studied via 
numerical integration of a time-dependent Schr\"{o}dinger equation on a grid approach. 
Single, double and triple ionization yields originating from a completely antisymmetric wave function are calculated and extracted using a restricted-geometry model with the soft-core potential and three active electrons.
The direct triple ionization channel is found to produce a larger yield than the channel connected with single and then direct double ionization.
Compared against earlier results investigating the $n s^ 2 n p^1$ configuration, we propose that the differences found here might in fact be accessible through electron's momentum distribution.
\end{abstract}

\maketitle

\section{\label{sec:introduction}Introduction}
The study of correlations is the study of the complexity of the world around us. One of the  
amazing manifestations of the existence of correlations in nature is the phenomenon of non-sequential double ionization (NSDI) in strong laser fields~\cite{Liu99, Bergues12}. 
Reports from experiments showing the recorded double ionization yield higher by several orders of magnitude than expected in the sequential electron escape processes~\cite{Fittinghoff92, Kondo93, walker1994precision}, followed by the measurements of the ion recoil momentum and latter extraction of electrons' momenta distributions with the famous finger-like structure~\cite{staudte2007binary, rudenko2007correlated} forced researchers to acknowledge the fundamental role of electron-electron correlations played in NSDI. 
Along with the experimental work, there were attempts to theoretically explain the observed phenomenon. It is now recognized that the process has a stepwise character and the rescattering is in focus. In short, one of the electrons tunnels and begins to move away from its parent ion. When the phase of the field changes (we deal with short pulses, usually having the wavelength on the border of visible and infrared light), the electron is turned back, accelerated and forced to recollide with the ion. As a result of the recollision, energy transfer occurs and consequently the escape of the second electron is allowed.

Higher ionization yield is also observed in the processes involving three or more electrons, then we speak of non-sequential multiple ionization (NSMI)~\cite{larochelle1998non}. Here too, rescattering plays an important role. However, theoretical analysis of events involving more than two electrons is very difficult. This is evidenced by the fact that full-size, i.e. taking into account all spatial dimensions for each electron, quantum calculations even for two electrons are still very rare~\cite{parker1998intense, parker2000time, Parker06,feist2008nonsequential, Hao14}. Simplified quantum models with a reduced number of dimensions are often used to  
overcome the numerical difficulty~\cite{lein2000intense,ruiz2006ab,staudte2007binary,prauzner2008quantum, prauzner2007time, eckhardt2010phase,chen2010double,Thiede18,Efimov18,Efimov19,mandrysz19,Efimov20}. And in the  case when three and more electrons are involved, classical or semi-classical calculations are dominant~\cite{Grobe94,sacha2001triple, Emmanouilidou06, Ho07, ho2006plane,zhou10, tang13}. 

We have recently shown that it is possible to construct a model with a reduced geometry that enables a study of triple ionization~\cite{Thiede18}. Importantly, the electronic configuration of the target atoms begins to play a significant role. In the case when two electrons are involved in the process, it is usually assumed that they have opposite spins and therefore the spatial part of the wave function is symmetrical. If one considers the two-electron problem with the antisymmetric spatial wave function (e.g. corresponding to the $^3S$ metastable state in He), the NSDI is expected to be strongly suppressed \cite{Eckhardt2008}. 
When three electrons are at play, there is no possibility that the spatial wave function is symmetrical. 
The electron configuration of the target atoms is reflected in the symmetry of the wave function under consideration. And so, for alkali metals with $n s^2 n p^1$ configurations we will have a spatial wave function which is partially antisymmetric, while for elements with $p^3$ configuration (e.g. nitrogen) we will have a completely antisymmetric function.
Importantly, due to symmetry properties of the ground state for atoms with the $n s^2 n p^1$ configuration it is not possible to reduce the problem of double ionization to the model of two active electrons~\cite{Efimov19}. In contrary, such a reduction is possible in the case of atoms with the $p^3$ configuration~\cite{Efimov20}.

In the previous work~\cite{Thiede18} we considered the triple ionization events in atoms with the $n s^2 n p^1$ configuration. In that case, the dominant triple ionization channel was the sequential escape for fields with amplitudes $F = 0.2$ and higher (in the following we use atomic units, unless otherwise stated). Channels associated with non-sequential escape, i.e. (i) three electrons at once, (ii) first one then two electrons and (iii) first two and then one electron, are important for fields with amplitudes less than $F = 0.2$. For the range of the analyzed field amplitudes, the process in which one electron is ionized first, and then two, plays a dominant role among the three mentioned paths of non-sequential escape.

In the present paper, we concentrate on the influence of the initial state symmetry on triple ionization. For this purpose, we analyze  triple ionization events for atoms with the $p^3$ configuration in the outer shell and compare the results  with the physics in the $n s^2 n p^1$ configuration.
The paper is structured as follows. Section II A describes briefly the dimensional reduction applied in the model and the involved parameters, while section II B the space-division method and extraction of fluxes allowing for calculating channel contributions in (multi)-electron ionization. In section III we present the main results and compare them to the results obtained in our previous work with different electron configuration. Section IV contains the conclusions.

\section{\label{sec:model}Model and methods}
\subsection{Model}
\begin{figure}[t]
\begin{center}
\includegraphics[width=0.4\textwidth]{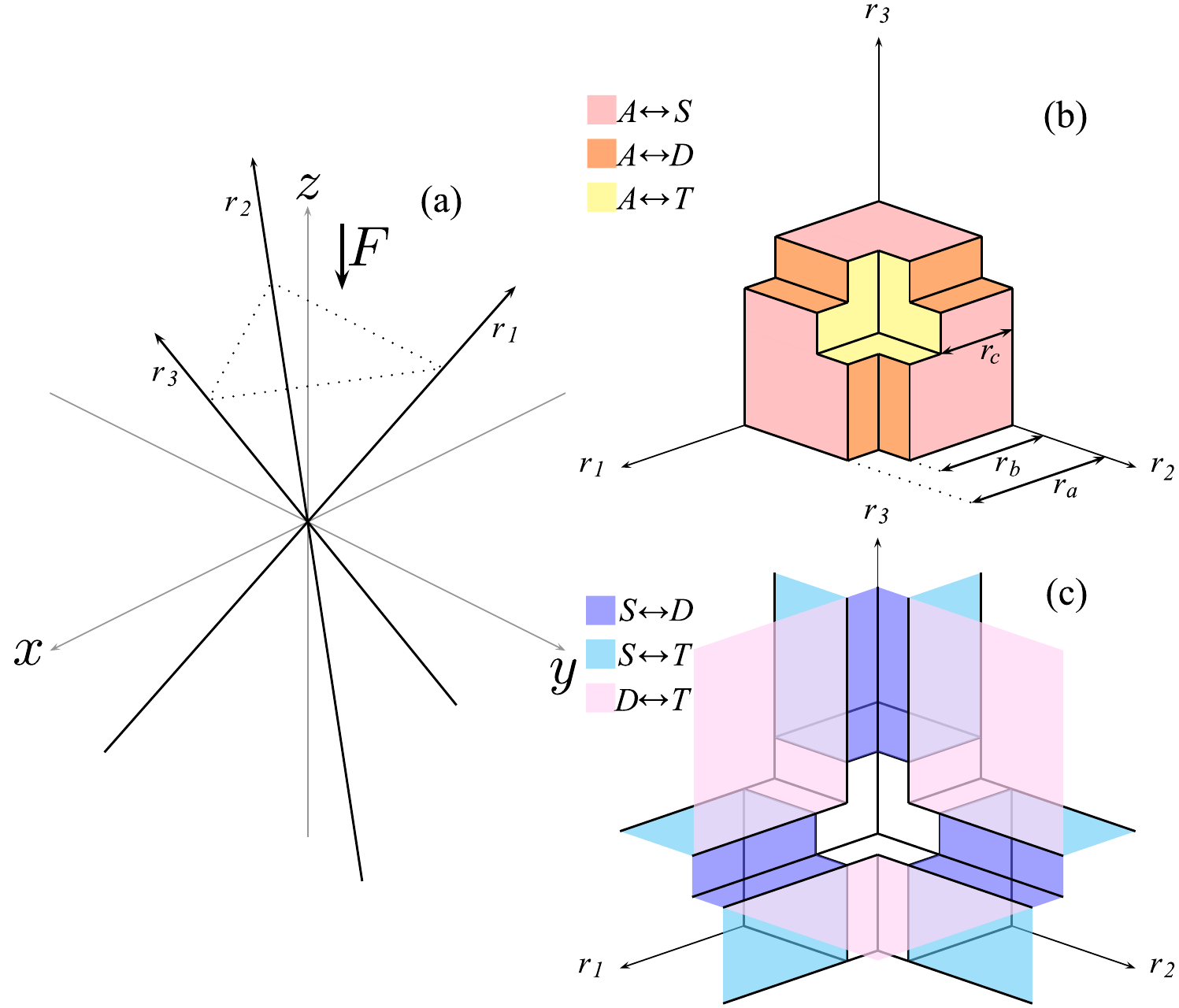}
\end{center}
\caption{The restricted-space model~\cite{Efimov20}.  Panel (a): The geometry of the model with respect to three-dimensional space: Electrons propagate along $r_1$, $r_2$, and $r_3$ axes. The field polarization direction, $\vec{F}$, is indicated by the arrow. Panels (b) and (c) - visualization of the space division within the model as used for calculation of probability fluxes. The space is divided into regions corresponding to neutral states ($A$), singly ionized states ($S$), doubly ionized states ($D$), and triply ionized states ($T$). The borders between respective regions are marked with different colors as shown in each panel: on panel (b) borders $A-S$, $A-D$ and $A-T$ are depicted, whereas on panel (c) borders $S-D$, $S-T$ and $D-T$.  The border distances are $r_a=12$ a.u., $r_b=7$ a.u., and $r_c=5$ a.u., respectively.}
\label{fig1}
\end{figure}

Due to a computational complexity it is virtually impossible nowadays to tackle the three electron problem numerically in the full phase space. Therefore, we employ a judiciously designed restricted-space model~\cite{Thiede18} in which each of the three electrons is allowed to move along one-dimensional (1D) track. The chosen 1D-tracks are equivalent to the lines along which the saddles, formed by the instantaneous electric field in the potential, move when the field amplitude is varied (see Fig.~\ref{fig1}(a)). The saddles and their motion were determined with the application of local stability analysis in the adiabatic potential~\cite{arnol2013mathematical, eckhardt06}. The Hamiltonian in the restricted-space reads (in atomic units):
\begin{equation}
H = \sum_{i=1}^3 \frac{p_i^2}{2} +V_a+V_{int},
\end{equation} 
where $V_a$ is the atomic potential:
\begin{equation}
V_a = -\sum_{i=1}^3 \frac{3}{\sqrt{r_i^2+\epsilon^2}} + \sum_{i,j=1;i<j}^3 \frac{q^2_{ee}}{\sqrt{(r_i-r_j)^2+r_ir_j+\epsilon^2}},
\end{equation}
reproducing the experimental single- and double-ionization potentials for a nitrogen atom with soft-core parameter $\epsilon=\sqrt{1.02}$ and effective electron-electron charges $q_{ee}=\sqrt{0.5}$ (same as in \cite{Efimov20}). Potential $V_{int}$ describes an interaction with the external field:
\begin{equation}
V_{int} = \sqrt{\frac{2}{3}}F(t)(r_1+r_2+r_3).
\label{int}
\end{equation}
$r_i$ and $p_i$ are the $i$'th electron's coordinate and conjugated momentum, respectively.
The field is defined via its vector potential, $F(t)=-\partial A/\partial t$, and is polarized along the $z$ axis in full space:
\begin{equation}
A(t) = \frac{F_0}{\omega_0}\sin^2\left(\frac{\pi t}{T_p}\right)\sin(\omega_0 t+\varphi), \;\; 0<t<T_p.
\end{equation}
Here $F_0$, $\omega_0$, $T_0=2\pi n_c/\omega_0$, $\varphi$ and $n_c$ are the field amplitude, the pulse frequency, the pulse length, the carrier-envelope phase and the number of cycles. In the following we set $\omega_0=0.06$ which corresponds to 760 nm of laser wavelength and the number of cycles, $n_c=5$. The field amplitude and the carrier-envelope phase are varied, although the behavior has been fairly consistent along the whole domain of the carrier-envelope phase hence only the results for $\varphi =0$ are  presented.
As the field is polarized along the $z$ axis in the full space it has to be projected onto $r_i$ tracks. The projection imposes the geometric factor $\sqrt{2/3}$ in \eqref{int}.

\subsection{Methods}
To calculate ionization yields we apply a space division method as commonly used before in both classical and quantum-mechanical calculations~\cite{dundas99, Efimov18} - compare Fig.~\ref{fig1}(b) and Fig.~\ref{fig1}(c). The total space is divided into regions corresponding to neutral states ($A$), singly ionized states ($S$), doubly ionized states ($D$) and triply ionized states ($T$). Region A extends up to $r_a=12$ a.u. from the origin of coordinate system in each direction establishing in this way a volume capable of enclosing the neutral atom wave function. Region $S$ is defined as sum of regions for which two electrons are close to the nucleus (less than $r_b=7$ a.u. from the origin) and the third one is far away. Region $D$ is defined as sum of regions for which only one electron is still close to the nucleus (less than $r_c=5$ a.u from the origin) and two other electrons are already far away. The last region, $T$, is defined as a sum of regions for which all electrons are far away from the origin. The populations of $A$, $S$, $D$ and $T$ states are calculated as integrated probability fluxes through the borders of respective regions. The assignment of the regions is to some extent arbitrary and the position of the borders affects quantitatively the results. However, as verified before~\cite{Efimov18, Thiede18, Efimov19,Efimov20} the chosen values provide results that reflect the correct trends in the dynamics of the studied system.

A much simpler case of two-electron system is described in detail in~\cite{prauzner2008quantum,Efimov18}, here we just briefly  mention the methodology behind the calculations and present the borders $A-S$, $A-D$, and $A-T$ in Fig.~\ref{fig1}(b) and borders $S-D$, $S-T$, and $D-T$ Fig.~\ref{fig1}(c). The instantaneous value of the population in region $R$ is calculated via the integral:\
\begin{equation}
P_R({\bf r},t)=P_R({\bf r},0) - \int_0^t f_R (\tau){\rm d}\tau,
\end{equation}
where $f_R(\tau)$ represents probability flux over border of the R region, i.e.
\begin{equation}
f_R(\tau) = -\iint_{\partial R} {\bf j}({\bf r}, \tau) \cdot {\rm d}\boldsymbol{ \sigma}.
\end{equation}
Here ${\rm d}\boldsymbol{\sigma}$ is a surface element and $\partial R$ symbolizes border of the region $R$.

The space-division method allows straightforwardly to distinguish between direct and sequential escapes in case of double ionization. For instance, calculating the flux through $A-D$ border allows us to obtain ionization yield for direct double ionization, whereas calculations of the flux through $S-D$ border will give the ionization yield for the sequential process. In the case of triple ionization, the method allows only to separate direct escapes, i.e. through $A-T$ border, from non-direct escapes that are calculated via fluxes through $S-T$ and $D-T$ borders. The latter two fluxes represent processes that involve double ionization either as the second or the first step in a path leading to triple ion, respectively. At this stage, however, it is not possible to tell whether that double ionization being an intermediate step in the triple ionization is direct or sequential itself. We discuss this issue and its solution later in the text.

\section{\label{sec:results}Results and discussion}
\begin{figure}[!t]
\begin{center}
\includegraphics[width=0.5\textwidth]{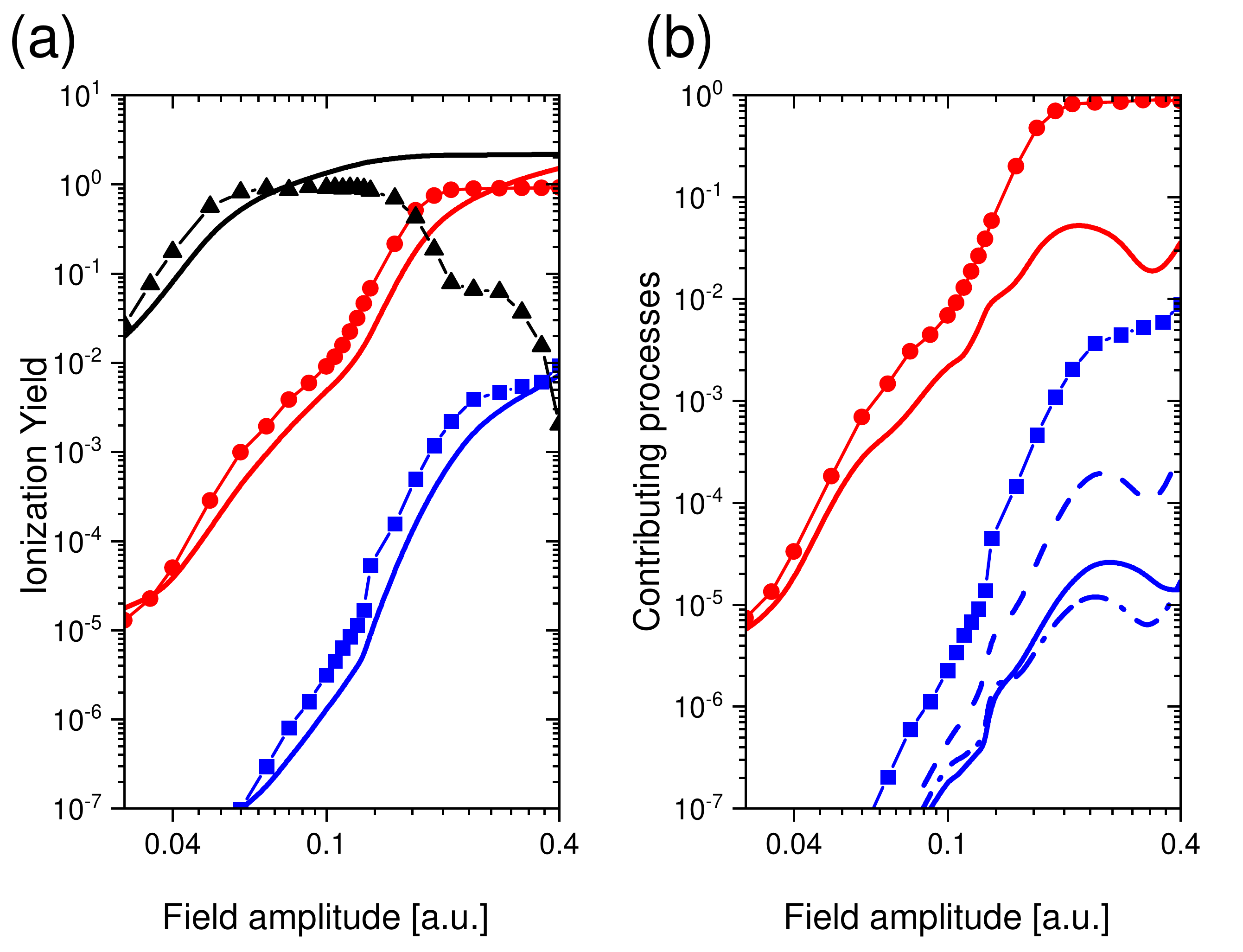}
\end{center}
\caption{Numerical ionization yields as a function of the peak electric field amplitude in atomic units. $F_0=0.1$ a.u. corresponds to $5.14\times 10^{10}$ V/m or laser intensity $I=3.5\times 10^{14}$ W/cm$^2$. 
Panel (a): Total yields for single ionization (black triangles), double ionization (red circles) and triple ionization (blue squares); solid lines represent averaged data after integration over a Gaussian beam. 
Panel (b): Different contributions to double and triple ionization - sets of red and blue lines, respectively. Solid lines without symbols show direct double (red) and triple (blue) escapes. Solid lines with symbols show sequential double (red line with circles) and sequential triple (blue line with squares) escapes. Dashed blue line presents double direct ionization followed by single ionization and dash-dotted blue line shows single ionization followed by direct double ionization.}
\label{fig2}
\end{figure}
\begin{figure}[t]
\begin{center}
\includegraphics[width=0.5\textwidth]{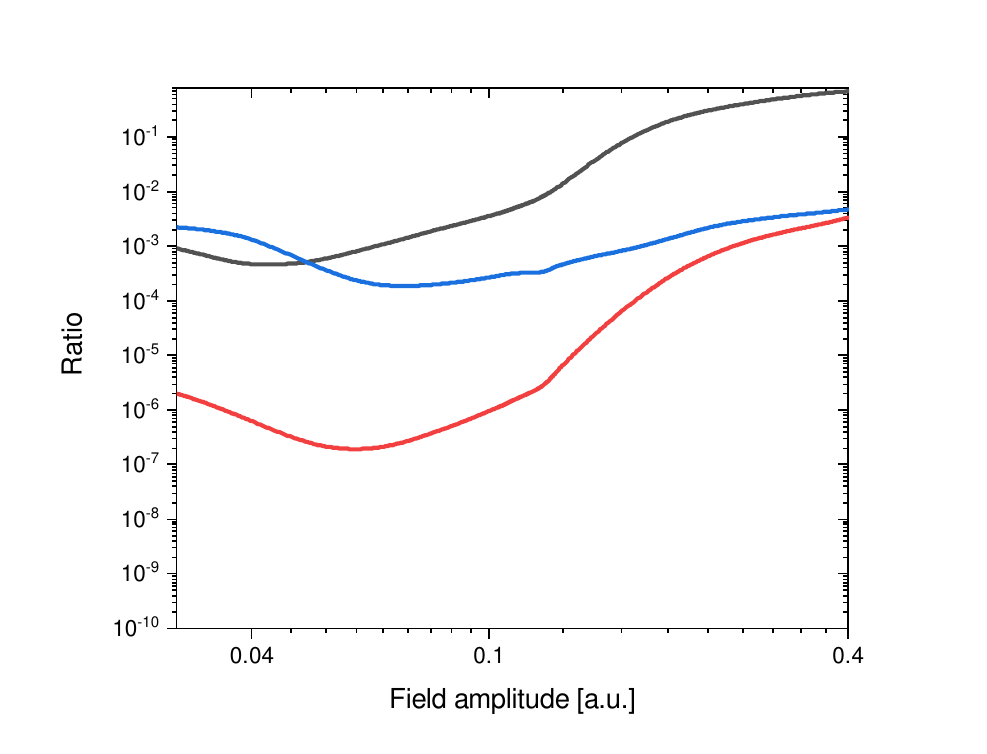}
\end{center}
\caption{Ratios of volume averaged numerical ionization yields as a function of peak electric field amplitude in atomic units: DI/SI ratio (black line), TI/SI (red line), and TI/DI (blue line).
}
\label{fig3}
\end{figure}

Let us first consider total ionization yields as a function of the peak electric field amplitude, see Fig.~\ref{fig2}(a). Results presented are obtained for the carrier-envelope phase $\varphi =0$. Single ionization (SI, black triangles) signal quickly saturates, then drops down for amplitudes larger than $F_0=0.1$ a.u. The observed drop of SI signal is a consequence of lack of averaging over intensity profile of the pulse which is inevitable in the experiments. Assuming a Gaussian profile of the laser beam the averaged ionization yield may be obtained as~\cite{strohaber15}:
\begin{equation}
P_{avg} (I_0) \propto \int_0^{I_0} \frac{P(I)}{I}{\rm d}I
\end{equation} 
The averaged SI yield is depicted with solid black line in Fig.~\ref{fig2} (a). As expected, once the saturation level is achieved it does not drop, because the higher intensity the lower weight is given to the respective yield. The same averaging procedure is used for double ionization (DI) and triple ionization (TI) yields and indicated by the corresponding solid lines.
Analyzing both the full DI yield (red circles) and its averaged counterpart (solid red line) in Fig.~\ref{fig2}(a) it is easy to notice the characteristic knee for amplitudes close to $F_0=0.1$ a.u. For larger field amplitude values DI signal still grows and then saturates eventually. Similar behavior is observed for TI yields, both non-averaged  (blue squares) and averaged (solid blue line). The only difference is that TI yield is three orders of magnitude smaller than DI yield, as may be inferred form Fig.~\ref{fig3} (see blue line) - similar observation was made in the case of $n s^2 n p^1$ electron configuration~\cite{Thiede18}. The ratio of double to single ionization yields (see black line in Fig.~\ref{fig3}) is practically constant for the field amplitude values corresponding to the characteristic knee in the yield curve and is of the order of $10^{-3}$ as reported in experiments \cite{walker1994precision, larochelle1998non}. Interestingly, similar dependence on the field amplitude is presented by the ratio of triple to single ionization yields (see red line in~Fig.~\ref{fig3}) - one could argue that there is a weak knee in TI yield too.

Multi-ionization signals may be further separated into different components due to the applied method of calculating yields (see Fig.~\ref{fig2}(b)). First, the double ionization signal is divided into two contributions, i.e. re-collision induced direct ionization (solid red line) and sequential ionization (red circles). The latter signal includes also contribution for re-collision excitation with a subsequent  ionization therefore the knee as a marker of re-collision importance is still visible. Sequential ionization signal rapidly grows for field amplitudes larger than $F_0=0.1$, while direct double ionization yield saturates at the level which is two orders of magnitude lower. Such a behavior is expected for  strong fields for which sequential ionization prevails over other processes.
	
It is interesting to look also at the different triple ionization contributions. The method of calculating ionization yields allows to straightforwardly differentiate triple ionization signal into three different contributions, namely, direct escape (0$\rightarrow$3), and two mixed paths, i.e. single ionization followed by double ionization and  double ionization followed by single ionization, where double ionization means any kind of double ionization, that is sequential and non-sequential. Therefore, these two mixed path both comprise some portion of process that is fully sequential (0$\rightarrow$1$\rightarrow$2$\rightarrow$3).
Such a situation, however, is of little interest as we would like to separate the sequential process from the direct one.

To resolve the above described difficulty we use an approach proposed by us while analyzing the case of $n s^2 n p^1$ electron configuration~\cite{Thiede18}, namely, we estimate different contributions to triple ionization based on what we learned about double ionization in this setup. More precisely, we assume that that the ratio of sequential to non-sequential double ionization, as determined by the fluxes through $S-D$ and $A-D$, holds for double ionization being the intermediate step in the three-electron process. Such an assumption allows us to extract the sequential contribution from the mixed paths. The results are presented in Fig.~\ref{fig2}(b) with a collection of blue lines. The signal that corresponds to a direct escape of three electrons is marked with the solid blue line, signals corresponding to partially direct escapes are marked with dashed and dash-dotted blue lines, and finally, signal for the sequential escape is marked with blue squares. As expected the sequential triple ionization (0$\rightarrow$1$\rightarrow$2$\rightarrow$3) dominates over the whole range of field amplitudes, however, other channels give non-negligible contributions too. Especially interesting is the fact that a direct triple ionization yield (0$\rightarrow$3) is not the lowest one. The weakest signal comes from the process in which single ionization is followed by a direct double ionization (0$\rightarrow$1$\rightarrow$3). The other partially direct process, i.e. a direct double ionization followed by  a single ionization (0$\rightarrow$2$\rightarrow$3) is much stronger than both (0$\rightarrow$3) and (0$\rightarrow$1$\rightarrow$3) pathways. The observed hierarchy of contributions is different from that obtained for Li-like atoms as reported earlier~\cite{Thiede18} for which direct escape was the least important channel of ionization. It is envisaged that such a difference in the hierarchy of contribution may influence electron's momentum distribution and thus be accessible in future experiments.

\section{\label{sec:conclusions}Conclusions}

We have studied triple ionization of atoms with $p^3$ valence shell. To this end we employed the restricted-geometry model with three active electrons solved on the grid. All obtained ionization yields feature trends observed in experiments, when the averaging that assumes the Gaussian profile of the laser beam is applied.
	With the application of the space-division method different contributions to triple ionization process are analyzed as a function of field amplitude values. The study reveals different hierarchy of contributions in atoms with $p^3$ valence shell as opposed to those with $n s^2 n p^1$. That difference is expected to have an influence on the electron's momentum distributions.

\section{Acknowledgements}
%TODO: Check if correct
We are grateful to Artur Maksymov for the help with the computer code. This work was supported by National Science Centre, Poland via Symfonia project No. 2016/20/W/ST4/00314 (MM, JPB and JZ). We also acknowledge the support of PL-Grid Infrastructure where all numerical calculations were carried out.

\bibliography{Prauzner-Bechcicki_p3}

\end{document}